\documentclass[11pt]{amsart}
\usepackage{amsmath,amsfonts,amscd,amssymb,epsf}
\newtheorem{theorem}{Theorem}[section]
\newtheorem{proposition}[theorem]{Proposition}

\theoremstyle{definition}

\theoremstyle{remark} \newtheorem{remark}[theorem]{Remark}
\numberwithin{equation}{section}

\renewcommand{\d}{\operatorname{d}}

\DeclareMathOperator{\dToda}{dToda}

\newcommand{\Z}{{\mathbb{Z}}}

\newcommand{\mL}{\mathcal{L}}

\newcommand{\F}{\mathcal{F}}

\newcommand{\B}{\mathcal{B}}

\newcommand{\pa}{\partial}

\begin{document}
\title[Toda Lattice Hierarchy]{Fay-like identities of Toda Lattice Hierarchy and its dispersionless limit}
\author{Lee-Peng Teo}\address{Faculty of Information
Technology, Multimedia University, Jalan Multimedia, Cyberjaya,
63100, Selangor Darul Ehsan, Malaysia}\email{lpteo@mmu.edu.my}

\begin{abstract}In this paper, we derive the Fay-like identities of
tau function for the Toda lattice hierarchy from the bilinear
identity. We prove that the Fay-like identities are equivalent to
the hierarchy. We also show that the dispersionless limit of the
Fay-like identities are the dispersionless Hirota equations of the
dispersionless Toda hierarchy.\end{abstract}

\maketitle

\section{Introduction}

The Toda lattice hierarchy was introduced   in \cite{UT} as a
generalization of Toda lattice (see e.g. \cite{Toda}). In the paper
\cite{UT}, Ueno and Takasaki developed the theory along the line of
the work of Date, Jimbo, Kashiwara and Miwa \cite{DJKM} on KP
hierarchy. In particular, they proved that there exists a tau
function for the Toda lattice hierarchy that satisfies a bilinear
identity, which implies one can consider KP hierarchy as a special
case of Toda lattice hierarchy.

In \cite{TT4}, Takasaki and Takebe considered the dispersionless
(quasi-classical) limit of the Toda lattice hierarchy. Since then,
the dispersionless Toda  (dToda) hierarchy has found to appear in a
lot of other areas of mathematics and physics, such as the evolution
of conformal mappings (see e.g. \cite{WZ, KKMWZ}), the solution of
Dirichlet boundary problem (see e.g. \cite{ref8}), WDVV equations
(see e.g. \cite{BMRWZ}), two-dimensional string theory (see e.g.
\cite{TT3}) and normal random matrix model (see e.g. \cite{WZ2}).
One of the ingredients appears in some of these works is the
dispersionless Hirota equations of the tau function of the dToda
hierarchy, first written down in \cite{WZ}, as analogues of the
dispersionless Hirota equation for dispersionless KP (dKP) hierarchy
 derived by Takasaki and Takebe in \cite{TT} (see also
 \cite{ref13_2}). In the Appendix of this seminal paper \cite{TT}, Takasaki and Takebe derived
the differential Fay identity from the bilinear identity satisfied
by the tau function of KP hierarchy. They showed that the
differential Fay identity is equivalent to KP hierarchy, and its
dispersionless limit is what we call dispersionless Hirota equation
of dKP hierarchy nowadays.
 However, up to date, we haven't found any
derivation of the dispersionless Hirota equation for dToda hierarchy
directly as dispersionless limits of  equations satisfied by the tau
function of Toda lattice hierarchy. The goal of the present paper to
solve this problem.

In Section \ref{Sec2}, we review some basic facts about the Toda
lattice hierarchy. In Section \ref{Sec3}, we re-derive the existence
of a tau function for Toda lattice hierarchy along the same line of
the proof of existence of tau function for KP hierarchy in
\cite{DJKM}. This section serves as a warm up for later sections. In
Section \ref{Sec4}, we derive what we call the Fay-like identities
for Toda lattice hierarchy from the bilinear identity satisfied by
the tau function. In Section \ref{Sec5}, we prove that the Fay-like
identities are equivalent to the Toda lattice hierarchy. More
specifically, a function satisfies the Fay-like identities if and
only if it is a tau function of the Toda lattice hierarchy. Finally,
in Section \ref{dispersionless}, we show that the dispersionless
limit of the Fay-like identities give the dispersionless Hirota
equations of dToda hierarchy.

\section{Toda Lattice Hierarchy}\label{Sec2}
In this section, we quickly review the necessary facts we need about
the Toda lattice hierarchy \cite{UT}.  We closely follow the
exposition in \cite{TT}.

Let $x=(x_1, x_2, \ldots)$ and $y=(y_1, y_2,\ldots)$ be two sets of
continuous variables. We denote by $s$ a continuous variable with
spacing unit $\hbar$. The  Lax formalism of Toda lattice hierarchy
is
\begin{align}\label{TodaLax}
\hbar\frac{\pa L}{\pa x_n}= [B_n, L], \hspace{1cm} \hbar\frac{\pa
L}{\pa y_n} =[C_n, L],\\
\hbar\frac{\pa K}{\pa x_n}= [B_n, K], \hspace{1cm} \hbar\frac{\pa
K}{\pa y_n} =[C_n, K],\nonumber
\end{align}where $L$, $K$, $B_n$, $C_n$ are   difference operators.   $L$ and $K^{-1}$ have the
form
\begin{align}\label{diffop}
L=& e^{\hbar\pa_s} +\sum_{n=0}^{\infty}
u_{n+1}^+(\hbar,s;x,y)e^{-n\hbar\pa_s},\\
K^{-1}=&u_0^-(\hbar,s;x,y)e^{-\hbar\pa_s}+\sum_{n=0}^{\infty}
u_{n+1}^-(\hbar,s;x,y)e^{n\hbar\pa_s},\nonumber
\end{align}
where the functions $u_n^{\pm}(\hbar,s;x,y)$ are assumed to be
regular in $\hbar$, i.e.,
$u_n^{\pm}(\hbar,s;x,y)=u_{n,0}^{\pm}(s;x,y)+O(\hbar)$ as
$\hbar\rightarrow 0$. $B_n$, $C_n$ are   defined by
\begin{align*}
B_n = (L^n)_{\geq 0},\hspace{1cm} C_n =(K^{-n})_{<0},
\end{align*}where for $A=\sum_{n\in \Z} A_n
e^{n \hbar\pa_s}$ a difference operator  and $S$ a subset of $\Z$,
we let $(A)_S=\sum_{n\in S} A_n e^{n\hbar\pa_s}$.

There exists two dressing operators $\hat{W}^{\pm}(\hbar, s; x,y)$
\begin{align*}
\hat{W}^{\pm}(\hbar,s;x,y)=\sum_{n=0}^{\infty} w_n^{\pm}(\hbar,s;
x,y) e^{\mp n\hbar\pa_s},\hspace{1cm} w_0^+(\hbar,s;x,y)\equiv 1,
\end{align*}
such that
\begin{align}\label{eq1}
L\hat{W}^+=& \hat{W}^+e^{\hbar\pa_s}, \hspace{3cm}
K\hat{W}^-=\hat{W}^-e^{\hbar\pa_s},\\
\hbar\frac{\pa \hat{W}^+ }{\pa x_n} =&-(L^n)_{<0}\hat{W}^+,
\hspace{1.8cm}\hbar\frac{\pa \hat{W}^+ }{\pa y_n}
=C_n\hat{W}^+,\nonumber\\
\hbar\frac{\pa \hat{W}^- }{\pa x_n} =&B_n\hat{W}^-,
\hspace{3cm}\hbar\frac{\pa \hat{W}^- }{\pa y_n} =-(K^{-n})_{\geq
0}\hat{W}^-.\nonumber
\end{align}
Let $\xi(t,a)=\sum_{n=1}^{\infty} t_n a^n$. Define
\begin{align}\label{eq5}
W^+(\hbar,
s;x,y)=&\hat{W}^+(\hbar,s;x,y)e^{\xi(x,e^{\hbar\pa_s})}=\left(\sum_{n=0}^{\infty}
w_n^{+}(\hbar,s;
x,y) e^{- n\hbar\pa_s}\right)e^{\xi(x,e^{\hbar\pa_s})},\\
W^-(\hbar,
s;x,y)=&\hat{W}^-(\hbar,s;x,y)e^{\xi(y,e^{-\hbar\pa_s})}=\left(\sum_{n=0}^{\infty}
w_n^{-}(\hbar,s; x,y) e^{
n\hbar\pa_s}\right)e^{\xi(y,e^{-\hbar\pa_s})}\nonumber.
\end{align}Then the system \eqref{eq1} is equivalent to
\begin{align*}
LW^+=& W^+e^{\hbar\pa_s}, \hspace{2cm}
KW^-=W^-e^{\hbar\pa_s},\\
\hbar\frac{\pa W^+ }{\pa x_n} =&B_nW^+, \hspace{2cm}\hbar\frac{\pa
W^+ }{\pa y_n}
=C_nW^+,\nonumber\\
\hbar\frac{\pa W^- }{\pa x_n} =&B_nW^-, \hspace{2cm}\hbar\frac{\pa
W^- }{\pa y_n} =C_nW^-.\nonumber
\end{align*}
Therefore
\begin{align}\label{eq6}
\hbar\frac{\pa W^+}{\pa x_n} \cdot(W^+)^{-1}=\hbar\frac{\pa W^-}{\pa
x_n} \cdot(W^-)^{-1},\\\hbar\frac{\pa W^+}{\pa y_n}\cdot
(W^+)^{-1}=\hbar\frac{\pa W^-}{\pa y_n}\cdot (W^-)^{-1}.\nonumber
\end{align}This gives the bilinear identity
\begin{align}\label{eq2}
W^+(\hbar, s; x,y)\cdot (W^+)^{-1}(\hbar,s;x',y')=W^-(\hbar, s;
x,y)\cdot (W^-)^{-1}(\hbar,s;x',y')
\end{align} for all $s, x,y,x',
y'.$ Let \begin{align*}
(\hat{W}^+)^{-1}(\hbar,s;x,y)=&\sum_{n=0}^{\infty} e^{-n\hbar\pa_s}
w_n^{+,*}(\hbar,s+\hbar;x,y),\\
(\hat{W}^-)^{-1}(\hbar,s;x,y)=&\sum_{n=0}^{\infty} e^{n\hbar\pa_s}
w_n^{-,*}(\hbar,s+\hbar;x,y),
\end{align*}and introduce the wave functions $w^{\pm} $ and the dual wave
functions $w^{\pm,*}$ by
\begin{align}\label{eq4}
w^{+}(\hbar, s;x,y;\lambda) = &\left(\sum_{n=0}^{\infty}
w_n^+(\hbar, s;x,y)\lambda^{-n}\right)\lambda^{s/\hbar}
e^{\xi(x,\lambda)/\hbar}\\=& \;\hat{w}^{+}(\hbar,
s;x,y;\lambda)\lambda^{s/\hbar}
e^{\xi(x,\lambda)/\hbar}\nonumber,\\
w^{+,*}(\hbar, s;x,y;\lambda) = &\left(\sum_{n=0}^{\infty}
w_n^{+,*}(\hbar, s;x,y)\lambda^{-n}\right)\lambda^{-s/\hbar}
e^{-\xi(x,\lambda)/\hbar}\nonumber\\=&\; \hat{w}^{+,*}(\hbar,
s;x,y;\lambda)\lambda^{-s/\hbar}
e^{-\xi(x,\lambda)/\hbar}\nonumber,\\
w^{-}(\hbar, s;x,y;\lambda) = &\left(\sum_{n=0}^{\infty}
w_n^-(\hbar, s;x,y)\lambda^{n}\right)\lambda^{s/\hbar}
e^{\xi(y,\lambda^{-1})/\hbar}\nonumber\\
=&\;\hat{w}^-(\hbar,s;x,y;\lambda)\lambda^{s/\hbar}
e^{\xi(y,\lambda^{-1})/\hbar}\nonumber,\\
w^{-,*}(\hbar, s;x,y;\lambda) = &\left(\sum_{n=0}^{\infty}
w_n^{-,*}(\hbar, s;x,y)\lambda^{n}\right)\lambda^{-s/\hbar}
e^{-\xi(y,\lambda^{-1})/\hbar}\nonumber\\
=&\;\hat{w}^{-,*}(\hbar,s;x,y;\lambda)\lambda^{-s/\hbar}
e^{-\xi(y,\lambda^{-1})/\hbar}.\nonumber
\end{align}
In terms of the wave functions and dual wave functions, the bilinear
identity \eqref{eq2} can be written in the residual form
\begin{align}\label{eq3}
&\text{Res}_{\lambda}\left(
w^+(\hbar,s;x,y;\lambda)w^{+,*}(\hbar,s';x',y';\lambda)\right)\\=&\text{Res}_{\lambda}\left(
w^-(\hbar,s;x,y;\lambda^{-1})w^{-,*}(\hbar,s';x',y';\lambda^{-1})\lambda^{-2}\right)\nonumber
\end{align}for all $s, s', x,x',y,y'$. Here for a power series $\beta(\lambda)=\sum_{n\in \Z}
\beta_n\lambda^n$,
$\text{Res}_{\lambda}(\beta(\lambda))=\beta_{-1}$.

On the other hand, we have
\begin{proposition}\label{pro1}
If $W^+$ and $W^-$ of the form \eqref{eq5} satisfy the system of
equations \eqref{eq6}, then
\begin{align*}
\hbar\frac{\pa W^+ }{\pa x_n} =&B_nW^+, \hspace{2cm}\hbar\frac{\pa
W^+ }{\pa y_n}
=C_nW^+,\nonumber\\
\hbar\frac{\pa W^- }{\pa x_n} =&B_nW^-, \hspace{2cm}\hbar\frac{\pa
W^- }{\pa y_n} =C_nW^-;\nonumber
\end{align*}and $(L,K)$ defined by
\begin{align*}
L=W^+e^{\hbar\pa_s}(W^+)^{-1},\hspace{1cm}K=W^-e^{\hbar\pa_s}(W^-)^{-1},
\end{align*}is a solution of the Toda lattice hierarchy.

\end{proposition}
\begin{proof}
See the proof of Theorem 1.5  in \cite{UT}.
\end{proof}

\section{Existence of Tau function}\label{Sec3}
Using the bilinear identity \eqref{eq2}, Ueno and Takasaki \cite{TT}
proved that there exists a tau function $\tau(\hbar, s; x,y)$ such
that
\begin{align}\label{eq39}
\hat{w}^+(\hbar,s;x,y;\lambda)=&\frac{\tau(\hbar,s;x-\hbar[\lambda^{-1}],y)}{\tau(\hbar,s;x,y)}\\
\hat{w}^{+,*}(\hbar,s;x,y;\lambda)=&\frac{\tau(\hbar,s;x+\hbar[\lambda^{-1}],y)}{\tau(\hbar,s;x,y)}\nonumber\\
\hat{w}^{-}(\hbar,s;x,y;\lambda)=&\frac{\tau(\hbar,s+\hbar;x,y-\hbar[\lambda])}{\tau(\hbar,s;x,y)}\nonumber\\
\hat{w}^{-,*}(\hbar,s;x,y;\lambda)=&\frac{\tau(\hbar,s-\hbar;x,y+\hbar[\lambda])}{\tau(\hbar,s;x,y)}\nonumber.
\end{align}Here $[\lambda]=\left(\lambda, \tfrac{1}{2}\lambda^2,
\tfrac{1}{3}\lambda^3, \ldots\right)$ and $\hat{w}^{\pm}$,
$\hat{w}^{\pm,*}$ are defined in \eqref{eq4}. As $\hbar \rightarrow
0$, the function $\log \tau(\hbar, s;x,y)$ behaves as (see
\cite{TT})
\begin{align}\label{taubehave}
\log\tau(\hbar,s;x,y)=\hbar^{-2}\F(s;x,y)+O(\hbar^{-1})
\end{align}
for some function $\F(s;x,y)$.

 In
this section, we  recapitulate  the proof of \cite{UT} along the
line of the proof of existence of tau function for KP hierarchy
given by \cite{DJKM}. We define the operators
\begin{align}\label{OpG}G^+(\mu)=\exp\left(-\hbar\sum_{n=1}^{\infty}\frac{\mu^{-n}}{n}\frac{\pa}{\pa
x_n}\right),\hspace{1cm}G^-(\nu)=\exp\left(-\hbar\sum_{n=1}^{\infty}\frac{\nu^{n}}{n}\frac{\pa}{\pa
y_n}\right)
\end{align}
and
$$d^+=\sum_{n=1}^{\infty} dx_n \frac{\pa}{\pa x_n},\hspace{1cm}d^-=\sum_{n=1}^{\infty} dy_n \frac{\pa}{\pa
y_n},\hspace{1cm} d=d^++d^-.$$
 We are going to make use of the following identities:
\begin{align}\label{eq7}
&\frac{1}{(1-\lambda
\eta_1)(1-\lambda\eta_2)}=\frac{\eta_1}{\eta_1-\eta_2}\frac{1}{1-\lambda\eta_1}-\frac{\eta_2}{
\eta_1-\eta_2}\frac{1}{1-\lambda\eta_2},
\\&\text{Res}_{\lambda}\left(\left(\sum_{n=0}^{\infty} \alpha_n
\lambda^{-n}\right)\frac{1}{(1-\lambda
\zeta^{-1})}\right)=\zeta\sum_{n=1}^{\infty}\alpha_n
\zeta^{-n}.\nonumber
\end{align}
First, we define
\begin{align}
\label{eq12}A_n^+(\hbar;
s;x,y)=&\text{Res}_{\lambda}\left(\lambda^{n}\left(-\sum_{j=1}^{\infty}\lambda^{-j-1}\hbar\frac{\pa}{\pa
x_j}+\frac{\pa}{\pa\lambda}\right)\log
\hat{w}^+(\hbar,s;x,y;\lambda)\right),\\A^-_n(\hbar;
s;x,y)=&-\text{Res}_{\lambda}\left(\lambda^{-n}\left(\sum_{j=1}^{\infty}\lambda^{j-1}\hbar\frac{\pa}{\pa
y_j}+\frac{\pa}{\pa\lambda}\right)\log
\hat{w}^-(\hbar,s;x,y;\lambda)\right)\nonumber\\
\omega=&\hbar^{-1} \sum_{n=1}^{\infty} A_n^+ dx_n+\hbar^{-1}
\sum_{n=1}^{\infty} A_n^-dy_n=\omega^++\omega^-,\nonumber
\end{align}
and rewrite the bilinear identity \eqref{eq3} as
\begin{align}\label{eq-bilinear}
&\text{Res}_{\lambda}\left( \hat{w}^+(\hbar,s;x,y;\lambda)
\hat{w}^{+,*}(\hbar,s';x',y';\lambda)\lambda^{(s-s')/\hbar}e^{(\xi(x,\lambda)-\xi(x',\lambda))/\hbar}\right)\\
=&\text{Res}_{\lambda}\left( \hat{w}^-(\hbar,s;x,y;\lambda^{-1})
\hat{w}^{-,*}(\hbar,s';x',y';\lambda^{-1})\lambda^{-2+((s'-s)/\hbar)}
e^{(\xi(y,\lambda)-\xi(y',\lambda))/\hbar}\right).\nonumber
\end{align}We consider the following cases:\\
\\
\textbf{Case I.} $s'=s$, $x'=x-\hbar[\mu_1^{-1}]-\hbar[\mu_2^{-1}]$,
$y'=y$.

In this case, the bilinear identity \eqref{eq-bilinear} gives
\begin{align*}
&\text{Res}_{\lambda}\Biggl( \hat{w}^+(\hbar,
s;x,y;\lambda)\hat{w}^{+,*}(\hbar,s;x-\hbar[\mu_1^{-1}]-\hbar[\mu_2^{-1}],y;\lambda)\\
&\hspace{5cm}\times\frac{1}
{(1-\lambda\mu_1^{-1})(1-\lambda\mu_2^{-1})}\Biggr)=0.
\end{align*}Using the   formulas in \eqref{eq7},
this gives
\begin{align}\label{eq9}
&\hat{w}^+(\hbar, s;
x,y;\mu_1)G^+(\mu_1)G^+(\mu_2)\hat{w}^{+,*}(\hbar,s;x,y;\mu_1)\\=&\hat{w}^+(\hbar,
s;
x,y;\mu_2)G^+(\mu_1)G^+(\mu_2)\hat{w}^{+,*}(\hbar,s;x,y;\mu_2).\nonumber
\end{align}Setting $\mu_1=\lambda$ and putting $\mu_2^{-1}=0$, we
have
\begin{align}\label{eq8}
\hat{w}^+(\hbar, s;
x,y;\lambda)G^+(\lambda)\hat{w}^{+,*}(\hbar,s;x,y;\lambda)=1
\end{align}or equivalently,
\begin{align}\label{eq10}
G^+(\lambda)\hat{w}^{+,*}(\hbar,s;x,y;\lambda)=&\frac{1}{\hat{w}^+(\hbar,
s; x,y;\lambda)}.
\end{align}
Using this relation, \eqref{eq9} gives
\begin{align*}
\frac{\hat{w}^+(\hbar, s;
x,y;\lambda_1)}{G^+(\lambda_2)\hat{w}^+(\hbar, s;
x,y;\lambda_1)}=&\frac{\hat{w}^+(\hbar, s;
x,y;\lambda_2)}{G^+(\lambda_1)\hat{w}^+(\hbar, s; x,y;\lambda_2)}
\end{align*}or equivalently,
\begin{align}\label{eq21}
&\log\hat{w}^+(\hbar, s;
x,y;\lambda_1)-G^+(\lambda_2)\log\hat{w}^+(\hbar, s;
x,y;\lambda_1)\\=&\log\hat{w}^+(\hbar, s;
x,y;\lambda_2)-G^+(\lambda_1)\log\hat{w}^+(\hbar, s;
x,y;\lambda_2)\nonumber.
\end{align}\\

 \noindent \textbf{Case II.} $s'=s+\hbar$,
$x'=x-\hbar[\mu^{-1}] $, $y'=y-\hbar[\nu]$.

In this case, the bilinear identity \eqref{eq-bilinear} gives
\begin{align*}
&\text{Res}_{\lambda}\left( \hat{w}^+(\hbar,s;
x,y;\lambda)\hat{w}^{+,*}(\hbar,s+\hbar; x-\hbar[\mu^{-1}],
y-\hbar[\nu];\lambda)\lambda^{-1}\frac{1}{1-\lambda\mu^{-1}}\right)\\
=&\text{Res}_{\lambda}\left( \hat{w}^-(\hbar,s ;
x,y;\lambda^{-1})\hat{w}^{-,*}(\hbar,s+\hbar ; x-\hbar[\mu^{-1}],
y-\hbar[\nu];\lambda^{-1})\lambda^{-1}\frac{1}{1-\lambda\nu}\right).
\end{align*}Using the second formula in \eqref{eq7}, this gives
\begin{align}\label{eq14}
&\hat{w}^+(\hbar,s;
x,y;\mu)G^+(\mu)G^-(\nu)\hat{w}^{+,*}(\hbar,s+\hbar; x , y
;\mu)\\=&\hat{w}^-(\hbar,s ;
x,y;\nu)G^+(\mu)G^-(\nu)\hat{w}^{-,*}(\hbar,s +\hbar; x , y
;\nu).\nonumber
\end{align}
Setting $\mu^{-1}=0$ and $\nu=\lambda$, we obtain
\begin{align}\label{eq15}
\hat{w}^-(\hbar,s; x,y;\lambda) G^-(\lambda)\hat{w}^{-,*}(\hbar,s
+\hbar; x , y ;\lambda)=1.
\end{align}
Equivalently,
\begin{align}\label{eq16}
G^-(\lambda)\hat{w}^{-,*}(\hbar,s; x , y
;\lambda)=\frac{1}{\hat{w}^-(\hbar,s-\hbar; x , y;\lambda)}.
\end{align}
Substituting \eqref{eq16} and  \eqref{eq10} into
\eqref{eq14}, we have
\begin{align*}
&\frac{\hat{w}^+(\hbar,s;
x,y;\lambda_1)}{G^-(\lambda_2)\hat{w}^{+}(\hbar,s+\hbar ; x , y
;\lambda_1)}=\frac{\hat{w}^-(\hbar,s;
x,y;\lambda_2)}{G^+(\lambda_1)\hat{w}^-(\hbar,s; x,y;\lambda_2)},
\end{align*}or equivalently,
\begin{align}\label{eq19}
&\log \hat{w}^+(\hbar,s;
x,y;\lambda_1)-G^-(\lambda_2)\log\hat{w}^{+}(\hbar,s+\hbar ; x , y
;\lambda_1)\\=&\log\hat{w}^-(\hbar,s;
x,y;\lambda_2)-G^+(\lambda_1)\log\hat{w}^-(\hbar,s;
x,y;\lambda_2)\nonumber.
\end{align}\\

\noindent \textbf{Case III.} $s'=s+2\hbar$, $x'=x  $,
$y'=y-\hbar[\nu_1]-\hbar[\nu_2] $.

In this case, the bilinear identity \eqref{eq-bilinear}   gives
\begin{align*}
0=\text{Res}_{\lambda}\Biggl(
\hat{w}^-(\hbar,s;x,y;\lambda^{-1})\hat{w}^{-,*}(\hbar,s+2\hbar;x,y-\hbar[\nu_1]-\hbar[\nu_2];\lambda^{-1})
\\\hspace{5cm}\times
\frac{1}{(1-\lambda\nu_1)(1-\lambda\nu_2)}\Biggr).
\end{align*}Using the   formulas in \eqref{eq7}, this gives
\begin{align*}
&\hat{w}^-(\hbar,s;x,y;\nu_1)G^-(\nu_1)G^-(\nu_2)\hat{w}^{-,*}(\hbar,s+2\hbar;x,y ;\nu_1)\\
=&\hat{w}^-(\hbar,s;x,y;\nu_2)G^-(\nu_1)G^-(\nu_2)\hat{w}^{-,*}(\hbar,s+2\hbar;x,y
;\nu_2).
\end{align*}Using \eqref{eq16}, this gives
\begin{align*}
\frac{\hat{w}^-(\hbar,s;x,y;\lambda_1)}{G^-(\lambda_2)
\hat{w}^-(\hbar,s+\hbar;x,y;\lambda_1)}=\frac{\hat{w}^-(\hbar,s;x,y;\lambda_2)}{G^-(\lambda_1)
\hat{w}^-(\hbar,s+\hbar;x,y;\lambda_2)},
\end{align*}or equivalently,
\begin{align}\label{eq25}
&\log\hat{w}^-(\hbar,s;x,y;\lambda_1)-G^-(\lambda_2)
\hat{w}^-(\hbar,s+\hbar;x,y;\lambda_1)\\=&\log\hat{w}^-(\hbar,s;x,y;\lambda_2)-G^-(\lambda_1)
\log\hat{w}^-(\hbar,s+\hbar;x,y;\lambda_2)\nonumber.
\end{align}
\vspace{1cm}

Now, using the definition \eqref{eq12} of $\omega^+$ and $\omega^-$,
\eqref{eq21}, \eqref{eq19} and \eqref{eq25} give us respectively
\begin{align}\label{eq13}
&\omega^+(\hbar,s;x,y)-G^+(\lambda)\omega^+(\hbar,s;x,y)=-
d^+\log\hat{w}^+(\hbar,
s; x,y;\lambda),\\
&\omega^-(\hbar,s;x,y)-G^+(\lambda)\omega^-(\hbar,s;x,y)=-
d^-\log\hat{w}^+(\hbar,s;
x,y;\lambda),\nonumber\\
&\omega^+(\hbar,s;x,y)-G^-(\lambda)\omega^+(\hbar,s+\hbar ;x,y)=-
d^+\log \hat{w}^-(\hbar,
s; x,y;\lambda), \nonumber\\
&\omega^-(\hbar,s;x,y)-G^-(\lambda)\omega^-(\hbar,s+\hbar;x,y)=-
d^-\log \hat{w}^-(\hbar,s; x,y;\lambda). \nonumber
\end{align}The first two equations and the last two equations give
respectively
\begin{align}\label{eq24}
&\omega(\hbar,s;x,y)-G^+(\lambda)\omega(\hbar,s;x,y)=-
d\log\hat{w}^+(\hbar,
s; x,y;\lambda), \\
&\omega(\hbar,s;x,y)-G^-(\lambda)\omega(\hbar,s+\hbar ;x,y)=- d\log
\hat{w}^-(\hbar, s;
x,y;\lambda).  \nonumber\\
\nonumber\end{align} Setting $\lambda=0$ in the second equation, we
have
\begin{align}\label{eq36}
\omega(\hbar,s;x,y)-\omega(\hbar,s+\hbar;x,y)=-  d\log
w_0^-(\hbar,s;x,y).
\end{align} Applying $d$ again to both sides of the
equations in   \eqref{eq24} and equation \eqref{eq36} , we obtain
\begin{align}\label{eq35}
d\omega(\hbar, s; x,y)=&G^+(\lambda)
d\omega(\hbar,s;x,y),\\d\omega(\hbar, s; x,y)=&G^-(\lambda)
d\omega(\hbar,s+\hbar;x,y),\nonumber\\
d\omega(\hbar, s; x,y)=&d\omega(\hbar,s+\hbar;x,y).\nonumber
\end{align}This implies that
\begin{align}\label{eq46}
d\omega(\hbar, s; x,y) = \sum_{n\neq 0}\sum_{m\neq 0} a_{mn}(\hbar
)dt_m \wedge dt_n,
\end{align}
where $t_n=x_n$ if $n>0$ and $t_n=y_n$ if $n<0$; $a_{mn}(\hbar )$
are independent of $x$ , $y$ and $s$ and $a_{nm}=-a_{mn}$.
Therefore,
\begin{align}\label{eq48}
\omega(\hbar, s; x,y) = \sum_{n\neq 0}\left(\sum_{m\neq 0}
a_{mn}(\hbar )t_m \right) dt_n+dH(\hbar,s;x,y)
\end{align}for some function $H(\hbar,s;x,y)$. Substituting back
into the equations in \eqref{eq24} , we have
\begin{align*}
  d\log\hat{w}^+(\hbar,s;x,y;\lambda)=&
G^+(\lambda)dH(\hbar,s;x,y)-dH(\hbar,s;x,y)\\& - \sum_{n\neq
0}\left(\sum_{m=1}^{\infty}
a_{mn}(\hbar )\frac{\lambda^{-m}}{m}\right)dt_n,\nonumber\\
  d\log \hat{w}^-(\hbar,s;x,y;\lambda) =&
G^-(\lambda)dH(\hbar,s+\hbar;x,y)-dH(\hbar,s;x,y)\nonumber\\& -
\sum_{n\neq 0}\left(\sum_{m=1}^{\infty}
a_{-m,n}(\hbar;s)\frac{\lambda^{m}}{m}\right)dt_n.\nonumber
\end{align*}Therefore, for some functions $A(\hbar,s;\lambda)=\sum_{n=1}^{\infty}A_n(\hbar,s)\lambda^{-n}$
 and $B(\hbar,s;\lambda)=\sum_{n=0}^{\infty}B_n(\hbar,s)\lambda^n$ independent of $x,y$,
we have
\begin{align}\label{eq45}
 \log\hat{w}^+(\hbar,s;x,y;\lambda)=&
G^+(\lambda)H(\hbar,s;x,y)-H(\hbar,s;x,y)\\& - \sum_{n\neq
0}\left(\sum_{m=1}^{\infty}
a_{mn}(\hbar )\frac{\lambda^{-m}}{m}\right)t_n+A(\hbar,s;\lambda),\nonumber\\
 \log \hat{w}^-(\hbar,s;x,y;\lambda) =&
G^-(\lambda)H(\hbar,s+\hbar;x,y)-H(\hbar,s;x,y)\nonumber\\& -
\sum_{n\neq 0}\left(\sum_{m=1}^{\infty}
a_{-m,n}(\hbar;s)\frac{\lambda^{m}}{m}\right)t_n+B(\hbar,s;\lambda)\nonumber.
\end{align}Substituting these back into equations \eqref{eq21}, \eqref{eq19} and
\eqref{eq25}, we find that
\begin{align*}
& \sum_{n=1}^{\infty}\left(\sum_{m=1}^{\infty} a_{mn}(\hbar
)\frac{\lambda_1^{-m}}{m}\right)\frac{\lambda_2^{-n}}{n}=
\sum_{n=1}^{\infty}\left(\sum_{m=1}^{\infty} a_{mn}(\hbar
)\frac{\lambda_2^{-m}}{m}\right)\frac{\lambda_1^{-n}}{n},\\
&\sum_{n=1}^{\infty}\left(\sum_{m=1}^{\infty} a_{m,-n}(\hbar
)\frac{\lambda_1^{-m}}{m}\right)\frac{\lambda_2^{n}}{n}+A(\hbar,s;\lambda_1)-A(\hbar,s+\hbar;\lambda_1)
\\=&\sum_{n=1}^{\infty}\left(\sum_{m=1}^{\infty}
a_{-m,n}(\hbar
)\frac{\lambda_2^{m}}{m}\right)\frac{\lambda_1^{-n}}{n},\\
&\sum_{n=1}^{\infty}\left(\sum_{m=1}^{\infty} a_{-m,-n}(\hbar
)\frac{\lambda_1^{m}}{m}\right)\frac{\lambda_2^{n}}{n}=\sum_{n=1}^{\infty}\left(\sum_{m=1}^{\infty}
a_{-m,-n}(\hbar
)\frac{\lambda_2^{m}}{m}\right)\frac{\lambda_1^{n}}{n}.
\end{align*}Comparing coefficients on both sides, we find that
$a_{mn}(\hbar)=a_{nm}(\hbar)$ for all $n\neq 0,m\neq 0$. But since
by definition $a_{nm}(\hbar)=-a_{mn}(\hbar)$, we conclude that
$a_{mn}(\hbar)=0$ for all $m\neq 0,n\neq 0$. Hence from
\eqref{eq46}, we have $ d\omega(\hbar,s;x,y)=0$. Together with
equation \eqref{eq36}, we conclude that there exists a function
$\tau(\hbar,s;x,y)$ such that
\begin{align}\label{eq37}
\omega(\hbar,s;x,y)=  d\log\tau(\hbar,s;x,y),\hspace{1cm}
\frac{\tau(\hbar, s+\hbar;x,y)}{\tau(\hbar,s;x,y)}=
w_0^-(\hbar,s;x,y).
\end{align}
We call $\tau(\hbar,s;x,y)$ the tau function of Toda lattice
hierarchy. Compare with \eqref{eq48}, we can take $H=  \log\tau$.
Then \eqref{eq45} gives us
\begin{align*}
\log\hat{w}^+(\hbar,s;x,y;\lambda)=&
G^+(\lambda)\log\tau (\hbar,s;x,y)-\log\tau(\hbar,s;x,y)+ A(\hbar,s;\lambda),\nonumber\\
\log \hat{w}^-(\hbar,s;x,y;\lambda) =&
G^-(\lambda)\log\tau(\hbar,s+\hbar;x,y)-\log\tau(\hbar,s;x,y)+
B(\hbar,s;\lambda)\nonumber.
\end{align*}
Substituting this into the definition \eqref{eq12} of $\omega$ and
using \eqref{eq37}, we conclude that
$A(\hbar,s;\lambda)=B(\hbar,s;\lambda)=0$. This gives the first and
third equations in \eqref{eq39}. Equations \eqref{eq10} and
\eqref{eq16} then give the second and fourth equations in
\eqref{eq39}.

\section{ Fay-like Identities}\label{Sec4}

In this section, we derive the
 Fay-like identities for Toda lattice hierarchy from the bilinear identity.

Substituting the equations \eqref{eq39} into the bilinear identity
\eqref{eq-bilinear}, we obtain the bilinear identity of Toda lattice
hierarchy in terms of the tau function:
\begin{align}\label{eq40}
&\text{Res}_{\lambda}\left(\tau(\hbar, s; x-\hbar[\lambda^{-1}],
y)\tau(\hbar,s';x'+\hbar[\lambda^{-1}],y')\lambda^{(s-s')/\hbar}e^{(\xi(x,\lambda)-\xi(x',\lambda))/\hbar}\right)\\
=&\text{Res}_{\lambda}\Bigl(\tau(\hbar, s+\hbar; x,
y-\hbar[\lambda^{-1}])
\tau(\hbar,s'-\hbar;x',y'+\hbar[\lambda^{-1}])\nonumber\\
&\hspace{5cm}\times\lambda^{-2+((s'-s)/\hbar)}
e^{(\xi(y,\lambda)-\xi(y',\lambda))/\hbar}\Bigr).\nonumber
\end{align}
We consider the following cases. \\
\\
\noindent
 \textbf{Case I.} $s'=s+\hbar$,
$x'=x-\hbar[\mu_1^{-1}]-\hbar[\mu_2^{-1}]$, $y'=y$.

 In this case,
the bilinear identity \eqref{eq40} and formulas in \eqref{eq7} give
us
\begin{align}\label{eq41}
&\frac{\mu_1^{-1}}{\mu_1^{-1}-\mu_2^{-1}}\tau(\hbar,s;x-\hbar[\mu_1^{-1}],y)\tau(\hbar,s+\hbar;x-\hbar[\mu_2^{-1}],
y)\\
-&\frac{\mu_2^{-1}}{\mu_1^{-1}-\mu_2^{-1}}\tau(\hbar,s;x-\hbar[\mu_2^{-1}],y)\tau(\hbar,s+\hbar;x-\hbar[\mu_1^{-1}],
y)\nonumber\\
=\;\;&\tau(\hbar,s+\hbar;x
,y)\tau(\hbar,s;x-\hbar[\mu_1^{-1}]-\hbar[\mu_2^{-1}], y).\nonumber
\end{align}\\

\noindent
 \textbf{Case II.} $s'=s+\hbar$,
$x'=x$, $y'=y-\hbar[\nu_1]-\hbar[\nu_2]$.

 In this case,
the bilinear identity \eqref{eq40} and formulas in \eqref{eq7} give
us
\begin{align}\label{eq42}
&\tau(\hbar, s; x,y)\tau(\hbar, s+\hbar; x,
y-\hbar[\nu_1]-\hbar[\nu_2])\\
=\;\;&\frac{\nu_1}{\nu_1-\nu_2}\tau(\hbar,s+\hbar; x,
y-\hbar[\nu_1])\tau(\hbar, s; x, y-\hbar[\nu_2])\nonumber\\
-&\frac{\nu_2}{\nu_1-\nu_2}\tau(\hbar,s+\hbar; x,
y-\hbar[\nu_2])\tau(\hbar, s; x, y-\hbar[\nu_1])\nonumber.
\end{align}\\

\noindent
 \textbf{Case III.} $s'=s $,
$x'=x-\hbar[\mu^{-1}] $, $y'=y-\hbar[\nu]$.

 In this case,
the bilinear identity \eqref{eq40} and formulas in \eqref{eq7} give
us
\begin{align}\label{eq43}
&\mu\Bigl(\tau(\hbar,s;x-\hbar[\mu^{-1}],y)\tau(\hbar,s;x,y-\hbar[\nu])-\tau(\hbar;s;x,y)\tau(\hbar
,s;x-\hbar[\mu^{-1}], y-\hbar[\nu])\Bigr)\\
=&\nu \tau(\hbar,s+\hbar; x,
y-\hbar[\nu])\tau(\hbar,s-\hbar;x-\hbar[\mu^{-1}], y )\nonumber.
\end{align}

\vspace{1cm} Rearranging \eqref{eq41}, \eqref{eq42} and
\eqref{eq43}, we obtain
\begin{align}\label{Fay}
\text{(A)}\hspace{2cm}&\mu_1-\mu_2\frac{\tau(\hbar,s;x-\hbar[\mu_1^{-1}],y)\tau(\hbar,s+\hbar;x-\hbar[\mu_2^{-1}],
y)}{\tau(\hbar,s;x-\hbar[\mu_2^{-1}],y)\tau(\hbar,s+\hbar;x-\hbar[\mu_1^{-1}],
y)}\\=&(\mu_1-\mu_2)\frac{\tau(\hbar,s+\hbar;x
,y)\tau(\hbar,s;x-\hbar[\mu_1^{-1}]-\hbar[\mu_2^{-1}],
y)}{\tau(\hbar,s;x-\hbar[\mu_2^{-1}],y)\tau(\hbar,s+\hbar;x-\hbar[\mu_1^{-1}],
y)},\nonumber\\
\text{(B)}\hspace{2cm}&\nu_1-\nu_2\frac{\tau(\hbar,s+\hbar; x,
y-\hbar[\nu_2])\tau(\hbar, s; x,
y-\hbar[\nu_1])}{\tau(\hbar,s+\hbar; x,
y-\hbar[\nu_1])\tau(\hbar, s; x, y-\hbar[\nu_2])}\nonumber\\
=&(\nu_1-\nu_2)\frac{\tau(\hbar, s; x,y)\tau(\hbar, s+\hbar; x,
y-\hbar[\nu_1]-\hbar[\nu_2])}{\tau(\hbar,s+\hbar; x,
y-\hbar[\nu_1])\tau(\hbar, s; x, y-\hbar[\nu_2])},\nonumber\\
\text{(C)}\hspace{2cm}& \frac{\tau(\hbar;s;x,y)\tau(\hbar
,s;x-\hbar[\mu^{-1}],
y-\hbar[\nu])}{\tau(\hbar,s;x-\hbar[\mu^{-1}],y)\tau(\hbar,s;x,y-\hbar[\nu])}
\nonumber\\=&1-\frac{\nu}{\mu}\frac{\tau(\hbar,s+\hbar; x,
y-\hbar[\nu])\tau(\hbar,s-\hbar;x-\hbar[\mu^{-1}], y )}{
\tau(\hbar,s;x-\hbar[\mu^{-1}],y)\tau(\hbar,s;x,y-\hbar[\nu])}\nonumber.
\end{align}We are going to prove in next section that these three
identities alone are enough to imply that $\tau(\hbar,s;x,y)$ is a
tau function of the Toda lattice hierarchy. We are also going to
prove in Section \ref{dispersionless} that the dispersionless limit
of these identities are precisely the dispersionless Hirota
equations of dispersionless Toda (dToda) hierarchy
\cite{ref6_1,WZ,ref8,ref9,BMRWZ, Teo}. These should be compared to
the work of Takasaki and Takebe in the Appendix of \cite{TT}, where
they showed that the differential Fay identity of KP hierarchy is
equivalent to the KP hierarchy, and the dispersionless limit of the
differential Fay identity is the dispersionless Hirota equation of
KP hierarchy \cite{TT, ref13_1,ref13_2}. Therefore we call the
identities (A)--(C) in \eqref{Fay} the Fay-like identities of Toda
lattice hierarchy.

\begin{remark}
In \cite{ref9}, Zabrodin has written down an equivalent form of the
identities (A) and (B) for a sector of the Toda lattice hierarchy
where the variables $x_n, y_n , n\in \mathbb{N}$ are complex
variables satisfying $y_n=-\bar{x}_n$.
\end{remark}

\section{Equivalence of Fay-like identities to the Toda Lattice
hierarchy}\label{Sec5}

In this section, we are going to show that the Fay-like identities
\eqref{Fay} are equivalent to the Toda lattice hierarchy. More
precisely, we have

\begin{proposition}
If $\tau(\hbar,s;x,y)$ is a function that satisfies the Fay-like
identities \eqref{Fay}, then $\tau(\hbar,s;x,y)$ is a tau function
of the Toda lattice hierarchy.
\end{proposition}
\begin{proof}
Given the function $\tau(\hbar,s;x,y)$ that satisfies the Fay-like
identities \eqref{Fay}, we define the functions
\begin{align*} \tilde{w}^{+}(\hbar, s;
x,y;\lambda)=&\frac{\tau(\hbar,s;x-\hbar[\lambda^{-1}],y)}{\tau
(\hbar,s;x,y)}e^{\xi(x,\lambda)}\\=&\left(\sum_{n=0}^{\infty}
w_n^{+}(\hbar,s;x,y)\lambda^{-
n}\right)e^{\xi(x,\lambda)},\\
\tilde{w}^{-}(\hbar, s;
x,y;\lambda)=&\frac{\tau(\hbar,s+\hbar;x,y-\hbar[\lambda])}{\tau
(\hbar,s;x,y)}e^{\xi(y,\lambda^{-1})}\\=&\left(\sum_{n=0}^{\infty}
w_n^{-}(\hbar,s;x,y)\lambda^{
n}\right)e^{\xi(y,\lambda^{-1})},\nonumber\end{align*} and define
the operators $W^{\pm}(\hbar,s;x,y)$ by \eqref{eq5}, with
$w_n^{\pm}(\hbar,s;x,y)$ defined by the formulas above. We are going
to show that $W^{\pm}$ satisfy the system of equations \eqref{eq6}.
Then by Proposition \ref{pro1}, $W^{\pm}$ are the dressing operators
of Toda lattice hierarchy. This implies that $\tau(\hbar,s;x,y)$ is
a tau function of the Toda lattice hierarchy.

We define the functions $\mathsf{w}^{\pm}, v^{\pm}$ by
\begin{align}\label{eq50}
\mathsf{w}^+(\hbar,s;x,y;\lambda)=&\frac{\tilde{w}^+(\hbar,s;x,y;\lambda)}{w_0^-(\hbar,s;x,y)}=
\frac{\tau(\hbar,s;x-\hbar[\lambda^{-1}],y)}{\tau(\hbar,s+\hbar;x,y)}e^{\xi(x,\lambda)},\\
\mathsf{w}^-(\hbar,s;x,y;\lambda)=&\frac{\tilde{w}^-(\hbar,s;x,y;\lambda)}{w_0^-(\hbar,s;x,y)}=
\frac{\tau(\hbar,s+\hbar;x,y-\hbar[\lambda
])}{\tau(\hbar,s+\hbar;x,y)}e^{\xi(y,\lambda^{-1})},\nonumber\\
v^+(\hbar,s;x,y;\lambda)
=&\lambda^{-1}\frac{\mathsf{w}^+(\hbar,s;x,y;\lambda)}{\mathsf{w}^+(\hbar,s+\hbar;x,y;\lambda)}
=\sum_{n=1}^{\infty}v_{n}^+(\hbar,s;x,y)
\lambda^{-n},\nonumber\\
v^-(\hbar,s;x,y;\lambda) =&\lambda
\frac{\tilde{w}^-(\hbar,s;x,y;\lambda)}{\tilde{w}^-(\hbar,s-\hbar;x,y;\lambda)}
=\sum_{n=1}^{\infty}v_{n}^-(\hbar,s;x,y) \lambda^{n},\nonumber
\end{align}
and  the operators $X[\mu]$, $Y[\nu]$ by
\begin{align*}
X[\mu] = 1-G^+(\mu),\hspace{1cm}Y[\nu]=1-G^-(\nu),
\end{align*}
where the operators $G^{\pm}$ are defined by \eqref{OpG}. It is easy
to verify that
\begin{align}\label{eq51}
\hbar\sum_{n=1}^{\infty}\frac{\mu^{-n}}{n}\frac{\pa}{\pa
x_n}=\sum_{n=1}^{\infty}\frac{X[\mu]^n}{n},
\hspace{1cm}\hbar\sum_{n=1}^{\infty}\frac{\nu^{n}}{n}\frac{\pa}{\pa
y_n}=\sum_{n=1}^{\infty}\frac{Y[\nu]^n}{n}.
\end{align}
Applying $X[\mu]$ to $\mathsf{w}^{\pm}$  and $Y[\nu]$ to $\tilde{w}^{\pm}$,
 we obtain respectively the following results:\\

\noindent \textbf{I.}
\begin{align*}
&X[\mu]\mathsf{w}^+(\hbar,s;x,y;\lambda)\\=&\frac{\tau(\hbar,s;x-\hbar[\lambda^{-1}],y)}{\tau
(\hbar,s+\hbar;x,y)}e^{\xi(x,\lambda)}-\frac{\tau(\hbar,s;x-\hbar[\lambda^{-1}]-\hbar[\mu^{-1}],y)}{\tau
(\hbar,s+\hbar;x-\hbar[\mu^{-1}],y)}e^{\xi(x,\lambda)}\left(1-\frac{\lambda}{\mu}\right)\\
=&\mathsf{w}^+(\hbar,s;x,y;\lambda)\left(1-\frac{\mu-\lambda}{\mu}
\frac{\tau
(\hbar,s+\hbar;x,y)\tau(\hbar,s;x-\hbar[\lambda^{-1}]-\hbar[\mu^{-1}],y)}{\tau(\hbar,s;x-\hbar[\lambda^{-1}],y)\tau
(\hbar,s+\hbar;x-\hbar[\mu^{-1}],y)} \right).\end{align*} By (A) of
\eqref{Fay}, this is equal to \begin{align*}
&\frac{\lambda}{\mu}\mathsf{w}^+(\hbar,s;x,y;\lambda)\frac{\tau(\hbar,s+\hbar;x-\hbar[\lambda^{-1}],y)\tau
(\hbar,s
;x-\hbar[\mu^{-1}],y)}{\tau(\hbar,s;x-\hbar[\lambda^{-1}],y)\tau
(\hbar,s+\hbar;x-\hbar[\mu^{-1}],y)}\\
=&\frac{\lambda}{\mu}\mathsf{w}^+(\hbar,s;x,y;\lambda)\frac{\mathsf{w}^+(\hbar,s+\hbar;x,y;\lambda)\mathsf
{w}^+(\hbar,s;x,y;\mu)}{\mathsf{w}^+(\hbar,s+\hbar;x,y;\mu)\mathsf
{w}^+(\hbar,s;x,y;\lambda)}\\
=&v^+(\hbar,s;x,y;\mu) \lambda
\mathsf{w}^+(\hbar,s+\hbar;x,y;\lambda).
\end{align*}\\

\noindent \textbf{II.}
\begin{align*}
&X[\mu]\mathbf{w}^-(\hbar,s;x,y;\lambda)\\=&\frac{\tau(\hbar,s+\hbar;x,y-\hbar[\lambda
])}{\tau(\hbar,s+\hbar;x,y)}e^{\xi(y,\lambda^{-1})}-\frac{\tau(\hbar,s+\hbar;x-\hbar[\mu^{-1}],y-\hbar[\lambda
])}{\tau(\hbar,s+\hbar;x-\hbar[\mu^{-1}],y)}e^{\xi(y,\lambda^{-1})}\\
=&\mathbf{w}^-(\hbar,s;x,y;\lambda)\left(1-
\frac{\tau(\hbar,s+\hbar;x,y)\tau(\hbar,s+\hbar;x-\hbar[\mu^{-1}],y-\hbar[\lambda
])}{\tau(\hbar,s+\hbar;x,y-\hbar[\lambda
])\tau(\hbar,s+\hbar;x-\hbar[\mu^{-1}],y)}\right).
\end{align*}By (C) of \eqref{Fay}, this is equal to
\begin{align*}
&\frac{\lambda}{\mu}\mathbf{w}^-(\hbar,s;x,y;\lambda)\frac{\tau(\hbar,s+2\hbar;x,y-\hbar[\lambda
])\tau(\hbar,s
;x-\hbar[\mu^{-1}],y)}{\tau(\hbar,s+\hbar;x,y-\hbar[\lambda
])\tau(\hbar,s+\hbar;x-\hbar[\mu^{-1}],y)}\\
=&\frac{\lambda}{\mu}\mathbf{w}^-(\hbar,s;x,y;\lambda)\frac{\mathsf{w}^-(\hbar,s+\hbar;x,y;\lambda
)\mathsf{w}^+(\hbar,s ;x ,y;\mu)}{\mathsf{w}^-(\hbar,s ;x,y;\lambda
)\mathsf{w}^+(\hbar,s+\hbar;x ,y;\mu)}\\
=&v^+(\hbar,s;x,y;\mu)
\lambda\mathbf{w}^-(\hbar,s+\hbar;x,y;\lambda).
\end{align*}

\noindent \textbf{III.}
\begin{align*}
&Y[\nu]\tilde{w}^+(\hbar,s;x,y;\lambda)\\=&
\frac{\tau(\hbar,s;x-\hbar[\lambda^{-1}],y)}{\tau
(\hbar,s;x,y)}e^{\xi(x,\lambda)}-\frac{\tau(\hbar,s;x-\hbar[\lambda^{-1}],y-\hbar[\nu])}{\tau
(\hbar,s;x,y-\hbar[\nu])}e^{\xi(x,\lambda)}\\
=&\tilde{w}^+(\hbar,s;x,y;\lambda)\left(1-\frac{\tau
(\hbar,s;x,y)\tau(\hbar,s;x-\hbar[\lambda^{-1}],y-\hbar[\nu])}{
\tau(\hbar,s;x-\hbar[\lambda^{-1}],y)\tau (\hbar,s;x,y-\hbar[\nu])}
\right)\\
=&\frac{\nu}{\lambda}\tilde{w}^+(\hbar,s;x,y;\lambda)\frac{\tau(\hbar,s-\hbar;x-\hbar[\lambda^{-1}],y)\tau
(\hbar,s+\hbar;x,y-\hbar[\nu])}{\tau(\hbar,s;x-\hbar[\lambda^{-1}],y)\tau
(\hbar,s;x,y-\hbar[\nu])}\\
=&\frac{\nu}{\lambda}\tilde{w}^+(\hbar,s;x,y;\lambda)\frac{\tilde{w}^+(\hbar,s-\hbar;x
,y;\lambda)\tilde{w}^- (\hbar,s
;x,y;\nu)}{\tilde{w}^+(\hbar,s;x,y;\lambda)\tilde{w}^-
(\hbar,s-\hbar;x,y;\nu)}\\
=&v^-(\hbar,s; x,y;\nu)
\lambda^{-1}\tilde{w}^+(\hbar,s-\hbar;x,y;\lambda).
\end{align*}

\noindent \textbf{IV.}
\begin{align*}
&Y[\nu]\tilde{w}^-(\hbar,s;x,y;\lambda)\\=&
\frac{\tau(\hbar,s+\hbar;x,y-\hbar[\lambda])}{\tau
(\hbar,s;x,y)}e^{\xi(y,\lambda^{-1})}\\&-\frac{\tau(\hbar,s+\hbar;x,y-\hbar[\lambda]-\hbar[\nu])}{\tau
(\hbar,s;x,y-\hbar[\nu])}e^{\xi(y,\lambda^{-1})}\left(1-\frac{\nu}{\lambda}\right)\\
=&\tilde{w}^-(\hbar,s;x,y;\lambda)\left(1-\frac{\tau
(\hbar,s;x,y)\tau(\hbar,s+\hbar;x,y-\hbar[\lambda]-\hbar[\nu])}{
\tau(\hbar,s+\hbar;x,y-\hbar[\lambda])\tau (\hbar,s;x,y-\hbar[\nu])}
\frac{\lambda-\nu}{\lambda} \right).\end{align*}By (B) of
\eqref{Fay}, this is equal to
\begin{align*}
&\frac{\nu}{\lambda}\tilde{w}^-(\hbar,s;x,y;\lambda)\frac{\tau(\hbar,s;x,y-\hbar[\lambda])\tau
(\hbar,s+\hbar;x,y-\hbar[\nu])}{\tau(\hbar,s+\hbar;x,y-\hbar[\lambda])\tau
(\hbar,s;x,y-\hbar[\nu])}\\
=&\frac{\nu}{\lambda}\tilde{w}^-(\hbar,s;x,y;\lambda)\frac{\tilde{w}^-(\hbar,s-\hbar;x,y;\lambda)\tilde{w}^-
(\hbar,s ;x,y;\nu)}{\tilde{w}^-(\hbar,s ;x,y;\lambda)\tilde{w}^-
(\hbar,s-\hbar;x,y;\nu)}\\
=&v^-(\hbar,s; x,y;\nu)
\lambda^{-1}\tilde{w}^-(\hbar,s-\hbar;x,y;\lambda).
\end{align*}

\vspace{0.5cm} Applying $X[\mu]$ again to \textbf{I} and \textbf{II}
we have
\begin{align*}
&X[\mu]^2
\mathsf{w}^{\pm}(\hbar,s;x,y;\lambda)\\=&\Bigl(X[\mu]v^+(\hbar,s;x,y;\mu)\Bigr)(\lambda
e^{\hbar\pa_s}) \mathsf{w}^{\pm}(\hbar,s;x,y;\lambda)
\\&+\Bigl(G(\mu)v^+(\hbar,s;x,y;\mu)\Bigr)(\lambda e^{\hbar\pa_s})\Bigl(X[\mu]
\mathsf{w}^{\pm}(\hbar,s;x,y;\lambda)\Bigr)
\\
=&\Biggl(\Bigl(X[\mu]v^+(\hbar,s;x,y;\mu)\Bigr)(\lambda
e^{\hbar\pa_s})\\
&\hspace{0.6cm}+\Bigl(G(\mu)v^+(\hbar,s;x,y;\mu)\Bigr)v^+(\hbar,s+\hbar;x,y;\mu)(\lambda
e^{\hbar\pa_s})^2\Biggr)\mathsf{w}^{\pm}(\hbar,s;x,y;\lambda)\\
=&\left(q_{2,1}(\hbar,s;x,y;\mu)(\lambda
e^{\hbar\pa_s})+q_{2,2}(\hbar,s;x,y;\mu)(\lambda
e^{\hbar\pa_s})^2\right)\mathsf{w}^{\pm}(\hbar,s;x,y;\lambda) .
\end{align*}Since as $\mu\rightarrow \infty$,
$v^+(\hbar,s;x,y;\mu)= O(\mu^{-1})$, we have
\begin{align*}
&q_{2,1}(\hbar,s;x,y;\mu)=X[\mu] v^+(\hbar,s;x,y;\mu)=O(\mu^{-1}),\\
&q_{2,2}(\hbar,s;x,y;\mu)=
\Bigl(G(\mu)v^+(\hbar,s;x,y;\mu)\Bigr)v^+(\hbar,s+\hbar;x,y;\mu)=O(\mu^{-2}).
\end{align*}By induction, we find that
\begin{align*}
X[\mu]^n\mathsf{w}^{\pm}(\hbar; s;x,y;\lambda) =&
\sum_{k=1}^{n}q_{n,k}(\hbar,s;x,y;\mu)(\lambda
e^{\hbar\pa_s})^k\mathsf{w}^{\pm}(\hbar; s;x,y;\lambda),\end{align*}
where $q_{n,k}(\hbar,s;x,y;\mu)$ is a power series in $\mu^{-1}$,
containing $\mu^j$ with $j\leq -k$. Similarly, from \textbf{III} and
\textbf{IV}, we have
\begin{align*}
Y[\nu]^n\tilde{w}^{\pm}(\hbar; s;x,y;\lambda) =&
\sum_{k=1}^{n}r_{n,k}(\hbar,s;x,y;\nu)(\lambda^{-1}
e^{-\hbar\pa_s})^k\tilde{w}^{\pm}(\hbar; s;x,y;\lambda),
\end{align*}where $r_{n,k}(\hbar,s;x,y;\nu)$ is a power series in
$\nu$,  containing $\nu^j$ with $j\geq k$. Consequently, we obtain
from \eqref{eq51} that
\begin{align}\label{eq52}
\hbar\sum_{n=1}^{\infty}\frac{\mu^{-n}}{n}\frac{\pa
\mathsf{w}^{\pm}(\hbar,s;x,y;\lambda)}{\pa x_n}=&\sum_{n=1}^{\infty}
\frac{\mu^{-n}}{n} Q_n (\hbar,s;x,y; \lambda e^{\hbar\pa
s})\mathsf{w}^{\pm}(\hbar,s;x,y;\lambda),\\
\hbar\sum_{n=1}^{\infty}\frac{\nu^{n}}{n}\frac{\pa
\tilde{w}^{\pm}(\hbar,s;x,y;\lambda)}{\pa y_n}=&\sum_{n=1}^{\infty}
\frac{\nu^{n}}{n} R_n (\hbar,s;x,y; (\lambda e^{\hbar\pa
s})^{-1})\tilde{w}^{\pm}(\hbar,s;x,y;\lambda),\nonumber
\end{align}
where $Q_n (\hbar,s;x,y; \Lambda)$ and $R_n  (\hbar,s;x,y;\Lambda)$
are order $n$--polynomials in $\Lambda$ without constant term.
Define
$$\tilde{W}^{\pm} (\hbar,s;x,y)=
w_0^-(\hbar,s;x,y)^{-1}W^{\pm}(\hbar,s;x,y).$$ Using the definition
\eqref{eq5} of $W^{\pm}(\hbar,s;x,y)$, and the fact that for any
function $\mathcal{A}(\hbar,s;x,y)$
\begin{align*}\left.\Bigl(e^{\hbar\pa_s}
\left(\mathcal{A}(\hbar,s;x,y)e^{n\hbar\pa_s} \right)\Bigr)\right|_{
e^{\hbar\pa_s}=\lambda}=&\left.\left(\mathcal{A}(\hbar,s+\hbar;x,y)e^{(n+1)\hbar\pa_s}
\right)\right|_{ e^{\hbar\pa_s}=\lambda}\\=&(\lambda
e^{\hbar\pa_s})\left(\mathcal{A}(\hbar,s;x,y)\lambda^n
\right),\end{align*} we obtain by comparing coefficients in
\eqref{eq52} that
\begin{align*}
\hbar\frac{\pa \tilde{W}^{\pm} (\hbar,s;x,y)}{\pa x_n}=&
Q_n(\hbar,s;x,y;   e^{\hbar\pa s})\tilde{W}^{\pm} (\hbar,s;x,y)\\
\hbar\frac{\pa W^{\pm} (\hbar,s;x,y)}{\pa y_n}=& R_n(\hbar,s;x,y;
e^{\hbar\pa s})\tilde{W}^{\pm} (\hbar,s;x,y).
\end{align*}These give
\begin{align}\label{equ1}
\hbar\frac{\pa \tilde{W}^+}{\pa x_n}\cdot
(\tilde{W}^+)^{-1}=\hbar\frac{\pa \tilde{W}^-}{\pa x_n}\cdot
(\tilde{W}^-)^{-1},\\
\hbar\frac{\pa  W^+}{\pa y_n}\cdot (W^+)^{-1}=\hbar\frac{\pa
W^-}{\pa y_n}\cdot (W^-)^{-1}.\nonumber
\end{align}Finally, we have from the first equation,
\begin{align*}
\hbar\frac{\pa  W^+}{\pa x_n}\cdot (W^+)^{-1}=& \hbar\frac{\pa
w_0^-}{\pa x_n}(w_0^-)^{-1}+w_0^-\frac{\pa \tilde{W}^+}{\pa
x_n}\cdot (\tilde{W}^+ )^{-1}(w_0^-)^{-1}\\
=&\hbar\frac{\pa w_0^-}{\pa x_n}(w_0^-)^{-1}+w_0^-\frac{\pa
\tilde{W}^-}{\pa x_n}\cdot (\tilde{W}^-)^{-1} (w_0^-)^{-1}\\
=&\hbar\frac{\pa  W^-}{\pa x_n}\cdot (W^-)^{-1}.
\end{align*}Together with the second equation of \eqref{equ1} give
\eqref{eq6}. This concludes the proof.
\end{proof}

\section{Dispersionless limit of Fay-like
identities}\label{dispersionless}

In this section, we show that the dispersionless limit
($\hbar\rightarrow 0$) of the Fay-like identities \eqref{Fay} are
the dispersionless Hirota equations of dToda hierarchy. First we
review some facts about dToda hierarchy.

\subsection{Dispersionless Toda Hierarchy}
By taking the dispersionless limit ($\hbar\rightarrow 0$) of the
Toda lattice hierarchy \eqref{TodaLax}, we obtain the Lax
representation of the dToda hierarchy \cite{TT4, TT}:
\begin{align}\label{dToda}
\frac{\pa\mL}{\pa x_n} =\{ \B_n, \mL\}, \hspace{1cm} \frac{\pa
\mL}{\pa y_n}=\{\mathcal{C}_n,\mL\},\\
\frac{\pa \mathcal{K}}{\pa x_n}=\{ \B_n,\mathcal{K}\},\hspace{1cm}
\frac{\pa \mathcal{K}}{\pa
y_n}=\{\mathcal{C}_n,\mathcal{K}\}\nonumber.
\end{align}
Here $\mL=\mL(p;s;x,y)$ and $\mathcal{K}=\mathcal{K}(p;s;x;y)$ are
formal power series in the variable $p$, such that $\mL$ and
$\mathcal{K}^{-1}$ have the form
\begin{align*}
\mL=\mL(p;s;x,y)=& p+\sum_{n=0}^{\infty} u^{+}_{n+1,0}(s;x,y)p^{-n},\\
\mathcal{K}^{-1}=\mathcal{K}(p;s;x,y)^{-1}=&u^{-}_{0,
0}(s;x,y)p^{-1}+\sum_{n=0}^{\infty} u^{-}_{n+1, 0}(s;x,y)p^n,
\end{align*}
obtained by replacing the operator $e^{\hbar\pa_s}$ by $p$ and
taking the limit $\hbar\rightarrow 0$ in the definition of $L$ and
$K$ in \eqref{diffop}. $\B_n$ and $\mathcal{C}_n$ are defined by
\begin{align*}
\B_n = (\mL^n)_{\geq 0}, \hspace{1cm}\mathcal{C}_n=
(\mathcal{K}^{-n})_{<0},
\end{align*}
where now for a power series $\mathcal{A}=\sum_{n\in\Z}\mathcal{A}_n
p^n$ and a subset $S$ of $\Z$, we define $\mathcal{A}_S=\sum_{n\in
S} \mathcal{A}_n p^n$. $\{\,\cdot\,, \,\cdot\,\}$ is the Poisson
bracket $$\{f, g\}=p\frac{\pa f}{\pa p}\frac{\pa g}{\pa
s}-p\frac{\pa f}{\pa s}\frac{\pa g}{\pa p}$$ of dToda hierarchy. The
function $\F(s;x,y)$ defined by \eqref{taubehave} as the leading
term in the expansion of $\log\tau(\hbar,s;x,y)$ with respect to
$\hbar$, is called the free energy of the dToda hierarchy. The tau
function of dToda hierarchy $\tau_{\dToda}$ is then defined as
$\tau_{\dToda} =\exp \F$. It satisfies the dispersionless Hirota
equation of dToda hierarchy, which is the following set of equations
\cite{ref6_1,WZ,ref8,ref9,BMRWZ, Teo}:
\begin{align}
\label{dHirota1} &z_1 \exp \left(
-\sum_{n=1}^{\infty}\frac{1}{n}\frac{\pa^2\log\tau_{\dToda}}{\pa
s\pa x_n} z_1^{-n}\right)-z_2 \exp \left(
-\sum_{n=1}^{\infty}\frac{1}{n}\frac{\pa^2\log\tau_{\dToda}}{\pa
s\pa x_n} z_2^{-n}\right)\\\nonumber
=&(z_1-z_2)\exp\left(\sum_{m=1}^{\infty}\sum_{n=1}^{\infty}\frac{1}{mn}
\frac{\pa^2\log\tau_{\dToda}}{\pa x_m\pa x_n}z_1^{-m}z_2^{-n}\right),\\
\label{dHirota2}&
z_1^{-1}\exp\left(\sum_{n=1}^{\infty}\frac{1}{n}\frac{\pa^2\log\tau_{\dToda}}{\pa
s\pa y_n}z_1^n\right)-
z_2^{-1}\exp\left(\sum_{n=1}^{\infty}\frac{1}{n}\frac{\pa^2\log\tau_{\dToda}}{\pa
s\pa y_n}z_2^n\right)\\
\nonumber =&(z_1^{-1}-z_2^{-2})
\exp\left(\sum_{m=1}^{\infty}\sum_{n=1}^{\infty}\frac{1}{mn}
\frac{\pa^2\log\tau_{\dToda}}{\pa y_m\pa y_n}z_1^{m}z_2^{n}\right),\\
\label{dHirota3} & 1-z_1 z_2^{-1}\exp\left( \frac{\pa^2\F}{\pa
s^2}-\sum_{n=1}^{\infty}\frac{1}{n}\frac{\pa^2\log\tau_{\dToda}}{\pa
s\pa y_n} z_1^{n}
+\sum_{n=1}^{\infty}\frac{1}{n}\frac{\pa^2\log\tau_{\dToda}}{\pa
s\pa
x_n} z_2^{-n}\right)\\
\nonumber
=&\exp\left(\sum_{m=1}^{\infty}\sum_{n=1}^{\infty}\frac{1}{mn}
\frac{\pa^2\log\tau_{\dToda}}{\pa y_m\pa x_n}z_1^{m}z_2^{-n}\right).
\end{align}It was shown in \cite{BMRWZ,ref13_1,ref13_2,Teo}
that this set of equations is equivalent to the dToda hierarchy.

\subsection{Dispersionless limit of Fay-like identities }

Making use of the behavior of $\tau(\hbar,s;x,y)$ as
$\hbar\rightarrow 0$ given by \eqref{taubehave}, it is easy to show
that
\begin{proposition}
The dispersionless limits of the Fay-like identities (A)--(C) in
\eqref{Fay} are the equations \eqref{dHirota1}--\eqref{dHirota3}
respectively.
\end{proposition}
\begin{proof}
  Using the operator $G^{\pm}$ defined
by \eqref{OpG}, and by a suitable renaming of the variables, we
rewrite (A)--(C) in \eqref{Fay} as
\begin{align*}
\text{(A$^{\prime}$)}\hspace{1cm}&z_1-z_2 \exp\Biggl(
\Bigl(e^{\hbar\pa_s}-1\Bigr)\Bigl(G^+(z_2)-G^+(z_1)\Bigr)\log\tau(\hbar,s;x,y)\Biggr)\\
=&(z_1-z_2)\exp\Biggl(\Bigl(
e^{\hbar\pa_s}-G^+(z_2)\Bigr)\Bigl(1-G^+(z_1)\Bigr)\log\tau(\hbar,s;x,y)
\Biggr),\\
\text{(B$^{\prime}$)}\hspace{1cm}&z_2^{-1}-z_1^{-1}\exp\Biggl(\Bigl(e^{\hbar\pa_s}-1\Bigr)\Bigl(G^-
(z_2)-G^-(z_1)\Bigr)\log\tau(\hbar,s;x,y)\Biggr)\\
=&(z_2^{-1}-z_1^{-1})\exp\Biggl(\Bigl(1-e^{\hbar\pa_s}G^-(z_1)\Bigr)\Bigl(1-G^-(z_2)\Bigr)
\log\tau(\hbar,s;x,y)\Biggr),\\
\text{(C$^{\prime}$)}\hspace{1cm}&\exp\Biggl(
\Bigl(1-G^+(z_2)\Bigr)\Bigl(1-G^-(z_1)\Bigr)\log\tau(\hbar,s;x,y)\Biggr)\\
= & 1- z_1
z_2^{-1}\exp\Biggl(\Bigl(e^{\hbar\pa_s}-1\Bigr)\Bigl(G^-(z_1)-e^{-\hbar\pa_s}G^+(z_2)\Bigr)
\log\tau(\hbar,s;x,y)\Biggr).
\end{align*}Since as $\hbar\rightarrow 0$,
\begin{align*}
G^+(\mu) =& \exp\left(-\hbar\sum_{n=1}^{\infty}
\frac{\mu^{-n}}{n}\frac{\pa}{\pa x_n}\right)=
1-\hbar\sum_{n=1}^{\infty} \frac{\mu^{-n}}{n}\frac{\pa}{\pa
x_n}+O(\hbar^2),\\
G^-(\nu) =& \exp\left(-\hbar\sum_{n=1}^{\infty}
\frac{\nu^{n}}{n}\frac{\pa}{\pa y_n}\right)=
1-\hbar\sum_{n=1}^{\infty} \frac{\nu^{n}}{n}\frac{\pa}{\pa
y_n}+O(\hbar^2),
\end{align*}we have
\begin{align*}
&\Bigl(e^{\hbar\pa_s}-1\Bigr)\Bigl(G^+(z_2)-G^+(z_1)\Bigr)\\=&
\hbar^2 \frac{\pa}{\pa s} \left(\sum_{n=1}^{\infty}
\frac{z_1^{-n}}{n}\frac{\pa}{\pa x_n}-\sum_{n=1}^{\infty}
\frac{z_2^{-n}}{n}\frac{\pa}{\pa x_n}\right)+O(\hbar^3),\\
&\Bigl(
e^{\hbar\pa_s}-G^+(z_2)\Bigr)\Bigl(1-G^+(z_1)\Bigr)\\=&\hbar^2\left(\frac{\pa}{\pa
s}+\sum_{n=1}^{\infty}\frac{z_2^{-n}}{n}\frac{\pa}{\pa
x_n}\right)\sum_{n=1}^{\infty}\frac{z_1^{-n}}{n}\frac{\pa}{\pa
x_n}+O(\hbar^3),\end{align*}\begin{align*}
&\Bigl(e^{\hbar\pa_s}-1\Bigr)\Bigl( G^-(z_2)-G^-(z_1)\Bigr)= \hbar^2
\frac{\pa}{\pa s} \left(\sum_{n=1}^{\infty}
\frac{z_1^{n}}{n}\frac{\pa}{\pa y_n}-\sum_{n=1}^{\infty}
\frac{z_2^{n}}{n}\frac{\pa}{\pa y_n}\right)+O(\hbar^3),\\
&\Bigl(1-e^{\hbar\pa_s}G^-(z_1)\Bigr)\Bigl(1-G^-(z_2)\Bigr)=\hbar^2\left(\sum_{n=1}^{\infty}
\frac{z_1^{n}}{n}\frac{\pa}{\pa y_n}-\frac{\pa}{\pa
s}\right)\sum_{n=1}^{\infty} \frac{z_2^{n}}{n}\frac{\pa}{\pa
y_n}+O(\hbar^3),\\
&\Bigl(1-G^+(z_2)\Bigr)\Bigl(1-G^-(z_1)\Bigr)=\sum_{n=1}^{\infty}
\frac{z_2^{-n}}{n}\frac{\pa}{\pa x_n} \sum_{n=1}^{\infty}
\frac{z_1^{n}}{n}\frac{\pa}{\pa y_n}+O(\hbar^3),\\
&\Bigl(e^{\hbar\pa_s}-1\Bigr)\Bigl(G^-(z_1)-e^{-\hbar\pa_s}G^+(z_2)\Bigr)\\=&\hbar^2\frac{\pa}{\pa
s}\left( \frac{\pa}{\pa s}-\sum_{n=1}^{\infty}
\frac{z_1^{n}}{n}\frac{\pa}{\pa y_n}+\sum_{n=1}^{\infty}
\frac{z_2^{-n}}{n}\frac{\pa}{\pa x_n}\right)+O(\hbar^3).
\end{align*}Therefore, using the behavior of $\tau(\hbar,s;x,y)$
given by \eqref{taubehave}, we find that as $\hbar\rightarrow 0$,
(A$^{\prime}$)---(C$^{\prime}$) give
\begin{align*}
\text{(A$^{\prime\prime}$)}\hspace{1cm}&z_1-z_2\exp\left(\sum_{n=1}^{\infty}\frac{1}{n}\frac{\pa^2\F}{\pa
s\pa x_n}z_1^{-n}-\sum_{n=1}^{\infty}\frac{1}{n}\frac{\pa^2\F}{\pa
s\pa x_n}z_2^{-n}\right) \\=&
\exp\left(\sum_{n=1}^{\infty}\frac{1}{n}\frac{\pa^2\F}{\pa s\pa
x_n}z_1^{-n}+\sum_{m=1}^{\infty}\sum_{n=1}^{\infty}\frac{1}{mn}\frac{\pa^2\F}{\pa
x_m\pa x_n}z_1^{-m}z_2^{-n}\right),\\
\text{(B$^{\prime\prime}$)}\hspace{1cm}&z_2^{-1}-z_1^{-1}\exp\left(\sum_{n=1}^{\infty}\frac{1}{n}\frac{\pa^2\F}{\pa
s\pa y_n}z_1^{n}-\sum_{n=1}^{\infty}\frac{1}{n}\frac{\pa^2\F}{\pa
s\pa y_n}z_2^{n}\right)\\
=&\exp\left(\sum_{m=1}^{\infty}\sum_{n=1}^{\infty}\frac{1}{mn}\frac{\pa^2\F}{\pa
y_m\pa
y_n}z_1^{m}z_2^{n}-\sum_{n=1}^{\infty}\frac{1}{n}\frac{\pa^2\F}{\pa
s\pa y_n}z_2^{n}\right),\\
\text{(C$^{\prime\prime}$)}\hspace{1cm}&\exp\left(\sum_{m=1}^{\infty}\sum_{n=1}^{\infty}\frac{1}{mn}\frac{\pa^2\F}{\pa
y_m\pa x_n}z_1^{m}z_2^{-n}\right)\\
=&1-z_1z_2^{-1}\exp\left(\frac{\pa^2\F}{\pa
s^2}-\sum_{n=1}^{\infty}\frac{1}{n} \frac{\pa^2\F}{\pa s\pa y_n}
z_1^n +\sum_{n=1}^{\infty}\frac{1}{n}\frac{\pa^2\F}{\pa s\pa
x_n}z_2^{-n}\right).
\end{align*}After simple manipulations, it it easy to see that this
set of equations give the equations \eqref{dHirota1},
\eqref{dHirota2} and \eqref{dHirota3} respectively.
\end{proof}

\vspace{0.5cm}

\noindent \textbf{Acknowledgments.} I am grateful for the
hospitality of Academia Sinica   of Taiwan during my visit, when
part of this work was done. Special thanks go to   Derchyi Wu who
had arranged the visit. I would also like to thank J.S. Chang, Y.T.
Chen, N.C. Lee, J.C. Shaw, M.H. Tu and D.C. Wu for the discussions
we had during my visit. This work is partially supported by MMU
Internal Funding PR/2006/0590.


\begin{thebibliography}{10}

\bibitem{BMRWZ}
A.~Boyarsky, A.~Marshakov, O.~Ruchayskiy, P.~Wiegmann, and
A.~Zabrodin,
  \emph{Associativity equations in dispersionless integrable hierarchies},
  Phys. Lett. B \textbf{515} (2001), no.~3-4, 483--492.

\bibitem{ref13_2}
R.~Carroll and Y.~Kodama, \emph{Solution of the dispersionless
{H}irota
  equations}, J. Phys. A \textbf{28} (1995), no.~22, 6373--6387.

\bibitem{DJKM}
Etsur{\=o} Date, Masaki Kashiwara, Michio Jimbo, and Tetsuji Miwa,
  \emph{Transformation groups for soliton equations}, Nonlinear integrable
  systems---classical theory and quantum theory (Kyoto, 1981), World Sci.
  Publishing, Singapore, 1983, pp.~39--119.

\bibitem{ref13_1}
John Gibbons and Yuji Kodama, \emph{Solving dispersionless {L}ax
equations},
  Singular limits of dispersive waves (Lyon, 1991), NATO Adv. Sci. Inst. Ser. B
  Phys., vol. 320, Plenum, New York, 1994, pp.~61--66.

\bibitem{KKMWZ}
I.~K. Kostov, I.~M. Krichever, M.~Mineev-Weinstein, P.~B. Wiegmann,
and
  A.~Zabrodin, \emph{The {$\tau$}-function for analytic curves}, Random matrix
  models and their applications, Math. Sci. Res. Inst. Publ., vol.~40,
  Cambridge Univ. Press, Cambridge, 2001, pp.~285--299.

\bibitem{ref8}
A.~Marshakov, P.~Wiegmann, and A.~Zabrodin, \emph{Integrable
structure of the
  {D}irichlet boundary problem in two dimensions}, Comm. Math. Phys.
  \textbf{227} (2002), no.~1, 131--153.

\bibitem{ref6_1}
M.~Mineev-Weinstein, P.B. Wiegmann, and A.~Zabrodin,
\emph{Integrable sructure
  of interface dynamics}, Phys.Rev.Lett. \textbf{84} (2000), 5106--5109.

\bibitem{TT3}
K.~Takasaki, \emph{Dispersionless {T}oda hierarchy and
two-dimensional string
  theory}, Comm. Math. Phys. \textbf{170} (1995), no.~1, 101--116.

\bibitem{TT4}
K.~Takasaki and T.~Takebe, \emph{Quasi-classical limit of {T}oda
hierarchy and
  {$W$}-infinity symmetries}, Lett. Math. Phys. \textbf{28} (1993), no.~3,
  165--176.

\bibitem{TT}
Kanehisa Takasaki and Takashi Takebe, \emph{Integrable hierarchies
and
  dispersionless limit}, Rev. Math. Phys. \textbf{7} (1995), no.~5, 743--808.

\bibitem{Teo}
Lee-Peng Teo, \emph{Analytic functions and integrable
  hierarchies---characterization of tau functions}, Lett. Math. Phys.
  \textbf{64} (2003), no.~1, 75--92.

\bibitem{Toda}
Morikazu Toda, \emph{Studies of a non-linear lattice}, Phys. Rep.
\textbf{18C}
  (1975), no.~1, 1--123.

\bibitem{UT}
Kimio Ueno and Kanehisa Takasaki, \emph{Toda lattice hierarchy},
Group
  representations and systems of differential equations (Tokyo, 1982), Adv.
  Stud. Pure Math., vol.~4, North-Holland, Amsterdam, 1984, pp.~1--95.

\bibitem{WZ2}
P.~Wiegmann and A.~Zabrodin, \emph{Large scale correlations in
normal
  non-{H}ermitian matrix ensembles}, J. Phys. A \textbf{36} (2003), no.~12,
  3411--3424.

\bibitem{WZ}
P.~B. Wiegmann and A.~Zabrodin, \emph{Conformal maps and integrable
  hierarchies}, Comm. Math. Phys. \textbf{213} (2000), no.~3, 523--538.

\bibitem{ref9}
A.~V. Zabrodin, \emph{The dispersionless limit of the {H}irota
equations in
  some problems of complex analysis}, Teoret. Mat. Fiz. \textbf{129} (2001),
  no.~2, 239--257.

\end{thebibliography}
\end{document}